\documentclass[a4paper,12pt]{article}
\usepackage{graphics}
\usepackage{graphicx}
\usepackage[caption=false]{subfig}
\usepackage{jheppub,shuffle,scalefnt}
\usepackage{epsfig,psfrag}
\usepackage[utf8]{inputenc}
\graphicspath{ {figs/} }

\newcommand\eps{\epsilon}

\allowdisplaybreaks

\title{A planar four-loop form factor and cusp anomalous dimension in QCD}
\author[a]{Johannes M.~Henn,}
\author[b]{Alexander V. Smirnov,}
\author[c]{Vladimir A. Smirnov,}
\author[d]{\\Matthias Steinhauser}
\affiliation[a]{PRISMA Cluster of Excellence, Johannes Gutenberg University, 55099 Mainz,
Germany}
\affiliation[b]{Research Computing Center, Moscow State University,
119991, Moscow, Russia}
\affiliation[c]{Skobeltsyn Institute of Nuclear Physics of Moscow State University,\\
119991, Moscow, Russia}
\affiliation[d]{ Institut f\"ur Theoretische Teilchenphysik,  Karlsruhe Institute of Technology (KIT),\\
76128 Karlsruhe, Germany}
\emailAdd{henn@uni-mainz.de}
\emailAdd{asmirnov80@gmail.com}
\emailAdd{smirnov@theory.sinp.msu.ru}
\emailAdd{matthias.steinhauser@kit.edu}
\preprint{  \parbox[t]{30mm}{ MITP/16-032 \\ TTP16-011}}

\abstract{ We compute the fermionic contribution to the photon-quark
  form factor to four-loop order in QCD in the planar limit in
  analytic form. From the divergent part of the latter the cusp and collinear
  anomalous dimensions are extracted. Results are also presented for
  the finite contribution. We briefly describe our method to compute
  all planar master integrals at four-loop order.  
}

\begin{document}
\maketitle
\flushbottom


\section{Introduction}

One of the important tasks of modern high-energy particle physics is the
development of new methods to compute quantum corrections to physical cross
sections. This is particularly important in the context of Quantum
Chromodynamics (QCD) where higher order corrections often have a significant
numerical impact.  In this article we provide the first
next-to-next-to-next-to-next-to-leading order (N$^4$LO) contribution to a
three-point function within QCD. 
To be precise, we consider the photon-quark form factor, which is
a building block for N${}^{4}$LO cross sections. Namely, it is a
gauge-invariant part of virtual forth-order corrections for the
process $e^+e^-\to 2$~jets, or for Drell-Yan production at hadron colliders.

Denoting the photon-quark vertex function by $\Gamma^{\mu}_q$ the scalar form
factor is obtained via
\begin{eqnarray}
  F_q(q^2) &=& -\frac{1}{4(1-\epsilon)q^2}
  \mbox{Tr}\left( p_2\!\!\!\!\!/\,\,\, \Gamma^\mu_q p_1\!\!\!\!\!/\,\,\,
  \gamma_\mu\right)
  \,,
\end{eqnarray}
where $D=4-2\epsilon$ is the space-time dimension, $q=p_1+p_2$ and $p_1$
($p_2$) is the incoming (anti-)quark momentum.  We 
consider the large-$N_c$
expansion of $F_q(q^2)$.
As a consequence we only have
to consider the contributions of planar Feynman diagrams.

Results for $F_q$ can be used to probe the infrared structure of gauge
theories.  Form factors encapsulate universal infrared contributions
coming from soft exchanges between two partons. The general form of
the latter is known 
\cite{Mueller:1979ih,Collins:1980ih,Sen:1981sd,Magnea:1990zb,Korchemskaya:1992je,Sterman:2002qn} and depends on cusp and collinear anomalous
dimensions.

Two-loop corrections to $F_q$ have been computed more than 25 years
ago~\cite{Kramer:1986sg,Matsuura:1987wt,Matsuura:1988sm,Gehrmann:2005pd}.
The first three-loop result has been presented in
Ref.~\cite{Baikov:2009bg} and has later been confirmed in
Ref.~\cite{Gehrmann:2010ue}. Analytic results for the three-loop
form factor integrals were presented in
Ref.~\cite{Lee:2010ik}. In Ref.~\cite{Gehrmann:2010tu}, the results of
Ref.~\cite{Lee:2010ik}  have been used to compute $F_q$ at three
loops up to order $\epsilon^2$, i.e., transcendental weight eight, as
a preparation for the four-loop calculation.

\begin{figure}[t]
  \begin{center}
    \begin{tabular}{ccc}
      \includegraphics[width=.3\textwidth]{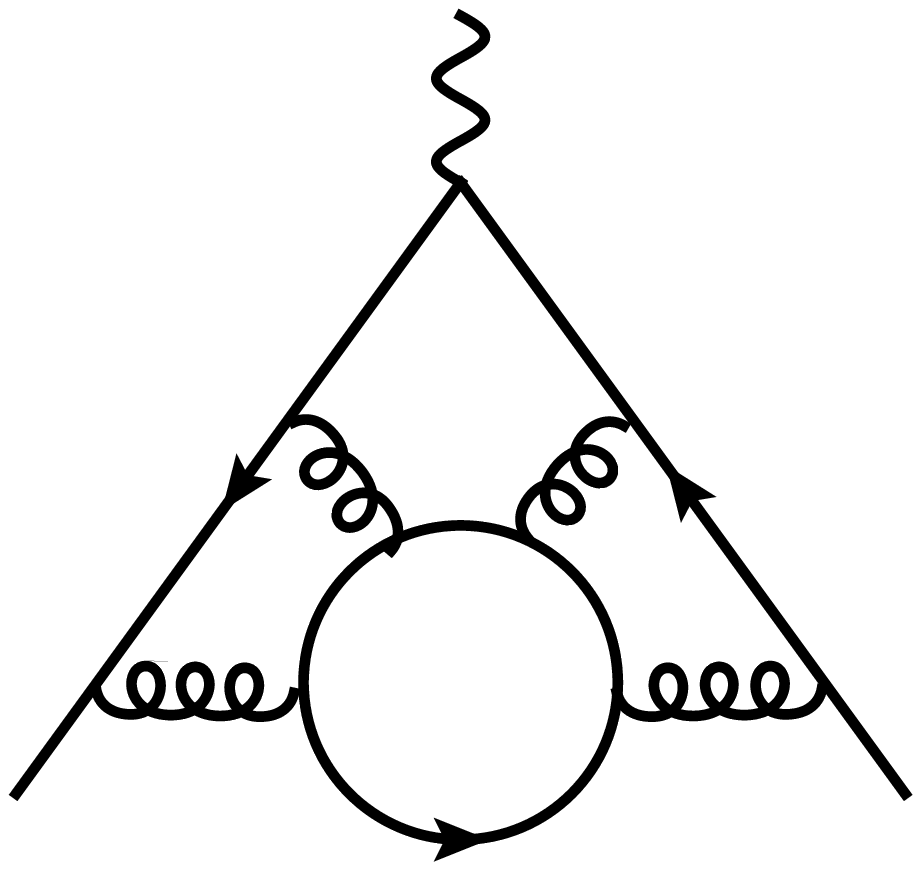} &
      \raisebox{.5em}
      {\includegraphics[width=.3\textwidth]{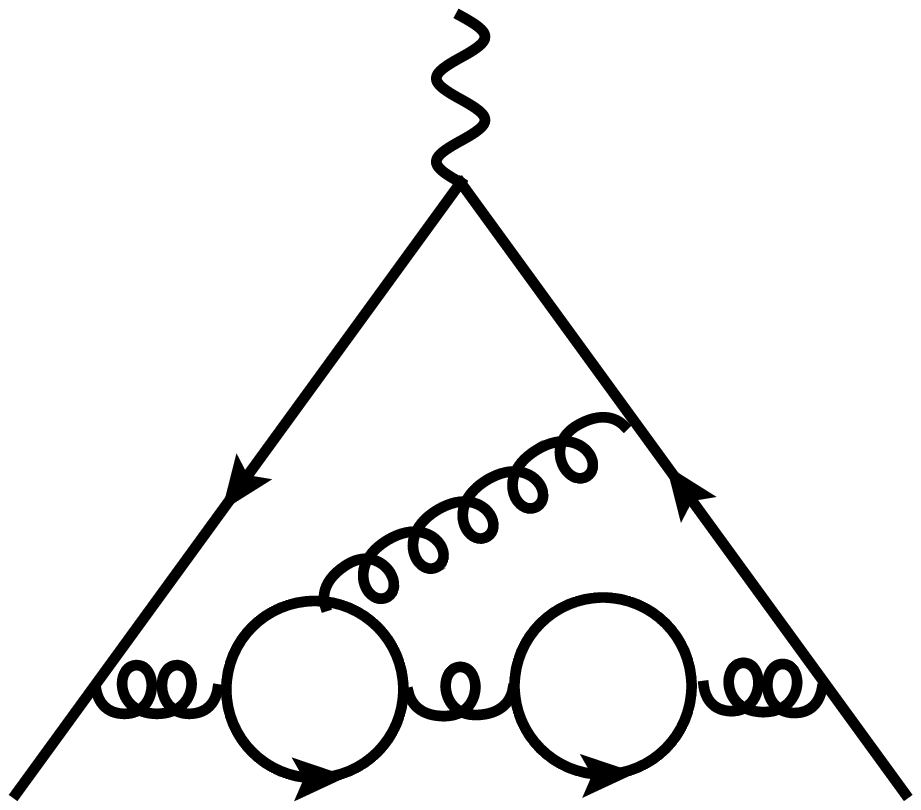}} &
      \includegraphics[width=.3\textwidth]{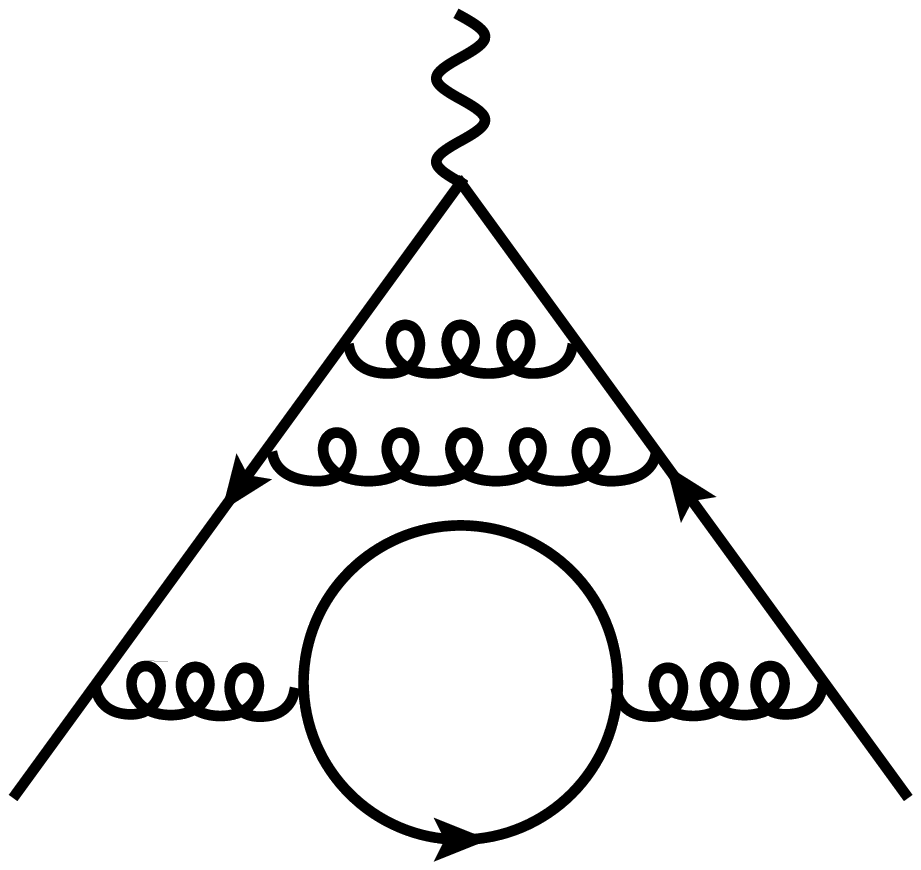}
    \end{tabular}
    \caption{\label{fig::diags}Sample Feynman diagrams contributing to
      the $F_q$ at four-loop order in the large-$N_c$ limit. Straight
      and curly lines denote quarks and gluons, respectively.
      The external wavy line represents the photon.}
  \end{center}
\end{figure}

In this paper we compute the fermionic corrections to $F_q$ in the
large-$N_c$ limit, to the four-loop order. 
Sample Feynman diagrams which have to be computed
for this purpose are shown in Fig.~\ref{fig::diags}.

Over the last decades, powerful methods for determining loop
integrands based on generalized unitarity have become common. However,
form factors are simple enough that a direct Feynman diagram approach
for determining the loop integrand is perfectly possible.  The
expression for the integrals contributing to the form factors is then
reduced to a set of so-called master integrals, exploiting
integration-by-parts identities \cite{Chetyrkin:1981qh}. This requires
rather involved computer algebra, and can be achieved using the latest
version of the program {\tt FIRE}~\cite{Smirnov:2008iw,Smirnov:2013dia,Smirnov:2014hma}.

This leaves the evaluation of the master integrals as the main technical
difficulty. In a previous paper \cite{Henn:2013nsa}, three of the present
authors proposed a new technique for computing such integrals.  Massless
form-factor integrals have a trivial scale dependence, so the powerful method
of differential
equations~\cite{Kotikov:1990kg,Bern:1993kr,Remiddi:1997ny,Gehrmann:1999as,Henn:2013pwa}
cannot be used directly. Rather, one first introduces an auxiliary parameter
(corresponding to a second off-shell external leg), in which differential
equations are set up. The main idea of~\cite{Henn:2013nsa} is that the
boundary value of the differential equations can be fixed trivially from a
value of the new parameter that corresponds to propagator-type integrals. This
boundary value is then related to the original problem via the differential
equations.

This last step is especially easy in the canonical form~\cite{Henn:2013pwa} of the
differential equations. It was suggested in that paper that in order to reach the 
canonical form it is helpful to select basis integrals that have constant leading singularities
\cite{Cachazo:2008vp}.  The latter are essentially multidimensional
residues of the loop integrand and can be computed
algorithmically. This connection makes it easy to reach the canonical
form of the differential equations, as was demonstrated in many recent papers.

We classified all massless planar four-loop form-factor integrals and
determined the corresponding master integrals. We found a total of
$99$ master integrals.  We then computed them as described in the
previous two paragraphs.  A very welcome by-product of the approach of
\cite{Henn:2013pwa} is that the results are typically expressed in
terms of uniform weight functions.  Examples of uniform weight form
factor integrals were previously considered in
Refs.~\cite{Lee:2010ik,Gehrmann:2011xn}.  Here we systematically found
a uniform weight basis for all planar integrals,
and expanded them to weight eight.
While this is the weight needed for typical four-loop computations, it is also 
possible to expand our result to higher weight.

Other attempts to calculate similar form factors or master integrals
were reported on in
Refs.~\cite{Boels:2012ew,Boels:2015yna,vonManteuffel:2015gxa}.
The evaluation of the master integrals in Refs.~\cite{Boels:2012ew,Boels:2015yna} was
performed only by numerical methods while Ref.~\cite{vonManteuffel:2015gxa}
presents results only for some individual integrals in an analytical form.

The remainder of the paper is structured as follows: In the next section we
briefly outline our calculation and present results for the
form factor and for the cusp and collinear anomalous dimensions.
Sections~\ref{sect:weight} and~\ref{sec::dgl} are dedicated to the
classification and evaluation of the master integrals. Our conclusions are
contained in Section~\ref{sec::concl}.


\section{\label{sec::calc_res}Calculation and results}

We generate the Feynman amplitudes with the help of {\tt
  qgraf}~\cite{Nogueira:1991ex} and transform the output to {\tt
  FORM}~\cite{Vermaseren:2000nd,Kuipers:2012rf} notation using {\tt
  q2e} and {\tt exp}~\cite{Harlander:1997zb,Seidensticker:1999bb}.
For the reduction to master integrals we use the program {\tt
  FIRE}~\cite{Smirnov:2008iw,Smirnov:2013dia,Smirnov:2014hma}  
which we apply in combination with {\tt LiteRed} \cite{Lee:2012cn,Lee:2013mka}.
Relations between primary master integrals occurring in the reduction
tables 
are revealed with the help of {\tt tsort}, which is part of the latest {\tt
  FIRE} version~\cite{Smirnov:2014hma}, and based on ideas presented in
Ref.~\cite{Smirnov:2013dia}. This leads to 78 master integrals  needed
for the fermionic part. More generally, we find that a total of 99 master
integrals are sufficient for arbitrary planar integrals. They are all computed
as described in Sections~\ref{sect:weight} and \ref{sec::dgl}.

In our calculation we allow for a generic QCD gauge parameter $\xi$
and expand the Feynman diagrams around $\xi=0$, which corresponds to Feynman
gauge, up to linear order. We checked that $\xi$ drops out before inserting
explicit results for the master integrals.
  
In the following we present results for the form factor $F_q$ and the
related anomalous dimensions. $F_q$ is conveniently shown in term of
the bare strong coupling constant. In that case the perturbative
expansion of $F_q$ can be cast in the form
\begin{eqnarray}
  F_q &=& 1 +
  \sum_{n\ge1} } 
  \left(\frac{\alpha_s^0}{4\pi}\right)^n
  \left(\frac{\mu^2}{-q^2} \right)^{(n\epsilon)}
  F_q^{(n)
  \,.
\end{eqnarray}
Analytic results for $F_q^{(n)}$, with $n\leq 3$, expanded in $\epsilon$ up to
transcendental weight eight can be found in
Ref.~\cite{Gehrmann:2010tu}. We refrain from repeating them here.

The main result of this letter is the fermionic contribution 
to $F_q^{(4)}$ in the large-$N_c$ limit. It is given by
\begin{eqnarray}
\lefteqn{  F_q^{(4)}|_{\mbox{\tiny large-$N_c$}} = } \nonumber\\
\nonumber\\&&\mbox{}
\frac{1}{\epsilon^7}  \Bigg[
\frac{1}{12} N_c^3 n_f
\Bigg]
%
%
+\frac{1}{\epsilon^6}  \Bigg[
\frac{41}{648} N_c^2 n_f^2-\frac{37}{648} N_c^3 n_f
\Bigg]
%
%
+\frac{1}{\epsilon^5}  \Bigg[
\frac{1}{54} N_c n_f^3 
+\frac{277}{972} N_c^2 n_f^2
\nonumber\\&&\mbox{}
+\left(\frac{41 \pi ^2}{648}-\frac{6431}{3888}\right) N_c^3 n_f
\Bigg]
%
%
+\frac{1}{\epsilon^4}  \Bigg[
\left(\frac{215 \zeta_3}{108}-\frac{72953}{7776}-\frac{227 \pi
  ^2}{972}\right) N_c^3 n_f
\nonumber\\&&\mbox{}
+\frac{11}{54} N_c n_f^3+\left(\frac{5}{24}+\frac{127 \pi ^2}{1944}\right) N_c^2 n_f^2
\Bigg]
%
%
+\frac{1}{\epsilon^3}  \Bigg[
\left(\frac{229 \zeta_3}{486}-\frac{630593}{69984}+\frac{293 \pi
  ^2}{2916}\right) N_c^2 n_f^2
\nonumber\\&&\mbox{}
+\left(\frac{2411 \zeta_3}{243}-\frac{1074359}{69984}-\frac{2125 \pi
  ^2}{1296}+\frac{413 \pi ^4}{3888}\right) N_c^3 n_f
+\left(\frac{127}{81}+\frac{5 \pi ^2}{162}\right) N_c n_f^3
\Bigg]
\nonumber\\&&\mbox{}
+\frac{1}{\epsilon^2}  \Bigg[
\left(-\frac{41 \zeta_3}{81}+\frac{29023}{2916}+\frac{55 \pi
  ^2}{162}\right) N_c n_f^3+\left(\frac{11684 \zeta_3}{729}-\frac{41264407}{419904}-\frac{155 \pi ^2}{72}
\right.\nonumber\\&&\left.\mbox{}
+\frac{2623 \pi ^4}{29160}\right) N_c^2 n_f^2+\left(-\frac{537625
  \zeta_3}{11664}-\frac{599 \pi ^2 \zeta_3}{486}+\frac{12853 \zeta_5}{180}+\frac{155932291}{839808}
\right.\nonumber\\&&\left.\mbox{}
-\frac{27377 \pi ^2}{69984}-\frac{1309 \pi ^4}{7290}\right) N_c^3 n_f
\Bigg]
%
%
+\frac{1}{\epsilon}  \Bigg[
\left(-\frac{451 \zeta_3}{81}+\frac{331889}{5832}+\frac{635 \pi
  ^2}{243}+\frac{151 \pi ^4}{4860}\right) N_c n_f^3
\nonumber\\&&\mbox{}
+\left(\frac{661
  \zeta_3}{4}-\frac{1805 \pi ^2 \zeta_3}{729}+\frac{19877
  \zeta_5}{405}-\frac{608092805}{839808}-\frac{6041473 \pi
  ^2}{209952}+\frac{8263 \pi ^4}{21870}\right) N_c^2
n_f^2
\nonumber\\&&\mbox{}
+\left(-\frac{5427821 \zeta_3}{5832}+\frac{48563 \pi ^2
  \zeta_3}{2916}-\frac{1373 \zeta_3^2}{324}+\frac{12847
  \zeta_5}{810}+\frac{662170621}{279936}+\frac{17271517 \pi
  ^2}{209952}
\right.\nonumber\\&&\left.\mbox{}
-\frac{78419 \pi ^4}{25920}+\frac{21625 \pi ^6}{81648}\right) N_c^3 n_f
\Bigg]
%
%
+ \Bigg[
\left(-\frac{10414 \zeta_3}{243}-\frac{205 \pi ^2
  \zeta_3}{243}-\frac{1097
  \zeta_5}{135}
+\frac{10739263}{34992}
\right.\nonumber\\&&\left.\mbox{}
+\frac{145115 \pi
  ^2}{8748}+\frac{1661 \pi ^4}{4860}\right) N_c
n_f^3+\left(\frac{65735207 \zeta_3}{52488}-\frac{4262 \pi ^2
  \zeta_3}{2187}-\frac{71711 \zeta_3^2}{1458}
\right.\nonumber\\&&\left.\mbox{}
+\frac{725828
  \zeta_5}{1215}-\frac{68487272627}{15116544}-\frac{295056623 \pi
  ^2}{1259712}-\frac{889 \pi ^4}{6480}+\frac{43559 \pi
  ^6}{204120}\right) N_c^2 n_f^2
\nonumber\\&&\mbox{}
+\left(-\frac{1774255975
  \zeta_3}{209952}+\frac{265217 \pi ^2 \zeta_3}{3888}-\frac{2692 \pi
  ^4 \zeta_3}{3645}+\frac{973135 \zeta_3^2}{1458}-\frac{56656921
  \zeta_5}{19440}
\right.\nonumber\\&&\left.\mbox{}
-\frac{58657 \pi ^2 \zeta_5}{1620}+\frac{1643545
  \zeta_7}{1008}+\frac{555003607961}{30233088}+\frac{785989381 \pi
  ^2}{839808}-\frac{34077673 \pi ^4}{2099520}
\right.\nonumber\\&&\left.\mbox{}
-\frac{146197 \pi
  ^6}{612360}\right) N_c^3 n_f
\Bigg] + \ldots
\,,
\end{eqnarray}
where the ellipses stand for $n_f$-independent contributions.

The cusp and collinear anomalous dimension is conveniently extracted
from $\log(F_q)$ (after renormalization of $\alpha_s$).
The pole part of the latter has the generic structure (see, e.g.,
Ref.~\cite{Gehrmann:2010ue,Becher:2009qa}) 
\begin{eqnarray}
\lefteqn{  \log( F_q )|_{\mbox{\tiny pole part}} =}
\nonumber\\&&\mbox{}
\frac{\alpha_s}{4\pi}  \Bigg\{
\frac{1}{\epsilon^2}  \Bigg[
-\frac{1}{2} C_F \gamma _{\text{cusp}}^0
\Bigg]
+\frac{1}{\epsilon}  \Bigg[
\gamma _q^0
\Bigg]
\Bigg\}
\nonumber\\&&\mbox{}
+\left(\frac{\alpha_s}{4\pi}\right)^2  \Bigg\{
\frac{1}{\epsilon^3}  \Bigg[
\frac{3}{8} \beta _0 C_F \gamma _{\text{cusp}}^0
\Bigg]
+\frac{1}{\epsilon^2}  \Bigg[
-\frac{1}{2} \beta _0 \gamma _q^0-\frac{1}{8} C_F \gamma _{\text{cusp}}^1
\Bigg]
+\frac{1}{\epsilon}  \Bigg[
\frac{\gamma _q^1}{2}
\Bigg]
\Bigg\}
\nonumber\\&&\mbox{}
+\left(\frac{\alpha_s}{4\pi}\right)^3  \Bigg\{
\frac{1}{\epsilon^4}  \Bigg[
-\frac{11}{36} \beta _0^2 C_F \gamma _{\text{cusp}}^0
\Bigg]
+\frac{1}{\epsilon^3}  \Bigg[
C_F \left(\frac{2}{9} \beta _1 \gamma _{\text{cusp}}^0+\frac{5}{36} \beta _0 \gamma _{\text{cusp}}^1\right)+\frac{1}{3} \beta _0^2 \gamma _q^0
\Bigg]
\nonumber\\&&\mbox{}
+\frac{1}{\epsilon^2}  \Bigg[
-\frac{1}{3} \beta _1 \gamma _q^0-\frac{1}{3} \beta _0 \gamma _q^1-\frac{1}{18} C_F \gamma _{\text{cusp}}^2
\Bigg]
+\frac{1}{\epsilon}  \Bigg[
\frac{\gamma _q^2}{3}
\Bigg]
\Bigg\}
\nonumber\\&&\mbox{}
+\left(\frac{\alpha_s}{4\pi}\right)^4  \Bigg\{
\frac{1}{\epsilon^5}  \Bigg[
\frac{25}{96} \beta _0^3 C_F \gamma _{\text{cusp}}^0
\Bigg]
+\frac{1}{\epsilon^4}  \Bigg[
C_F \left(-\frac{13}{96} \beta _0^2 \gamma _{\text{cusp}}^1-\frac{5}{12} \beta _1 \beta _0 \gamma _{\text{cusp}}^0\right)-\frac{1}{4} \beta _0^3 \gamma _q^0
\Bigg]
\nonumber\\&&\mbox{}
+\frac{1}{\epsilon^3}  \Bigg[
C_F \left(\frac{5}{32} \beta _2 \gamma _{\text{cusp}}^0+\frac{3}{32} \beta _1 \gamma _{\text{cusp}}^1+\frac{7}{96} \beta _0 \gamma _{\text{cusp}}^2\right)+\frac{1}{4} \beta _0^2 \gamma _q^1+\frac{1}{2} \beta _1 \beta _0 \gamma _q^0
\Bigg]
\nonumber\\&&\mbox{}
+\frac{1}{\epsilon^2}  \Bigg[
-\frac{1}{4} \beta _2 \gamma _q^0-\frac{1}{4} \beta _1 \gamma _q^1-\frac{1}{4} \beta _0 \gamma _q^2-\frac{1}{32} C_F \gamma _{\text{cusp}}^3
\Bigg]
+\frac{1}{\epsilon}  \Bigg[
\frac{\gamma _q^3}{4}
\Bigg]
\Bigg\}
\,,
  \label{eq::logFq_gen}
\end{eqnarray}
where $\mu^2=-q^2$ has been chosen and 
the coefficients of the $\beta$ function are given by
\begin{eqnarray}
  \beta_0 &=& \frac{11 C_A}{3}-\frac{2 n_f}{3}
  \,,\nonumber\\
  \beta_1 &=& -\frac{10 C_A n_f}{3}+\frac{34 C_A^2}{3}-2 C_F n_f
  \,,\nonumber\\
  \beta_2 &=& 
   -\frac{205}{18} C_A C_F n_f-\frac{1415}{54} C_A^2 n_f+\frac{79}{54}   C_A
    n_f^2+\frac{2857 C_A^3}{54}+\frac{11}{9} C_F n_f^2+C_F^2 n_f
  \,.
  \nonumber\\
\end{eqnarray}
The coefficients of the cusp and collinear anomalous dimensions
are defined through
\begin{eqnarray}
  \gamma_x &=& \sum_{n\ge0} \left(\frac{\alpha_s{(\mu^2)}}{4\pi}\right)^n
  \gamma_x^n
  \,,
  \label{eq::gamma_x}
\end{eqnarray}
with $x\in\{\text{cusp},q\}$.

From Eq.~(\ref{eq::logFq_gen}) it is evident that 
$\gamma_{\rm cusp}$ can be extracted from the coefficient of the
quadratic, and $\gamma_{q}$ from the first-order pole in $\eps$. In the
large-$N_c$ limit we obtain for $\gamma^{\rm cusp}$
\begin{eqnarray}
  \gamma_{\rm cusp}^0 &=& 4
  \,,\nonumber\\
  \gamma_{\rm cusp}^1 &=& \left(-\frac{4 \pi^2}{3}+\frac{268}{9}\right) N_c -\frac{40 n_f}{9}
  \,,\nonumber\\
  \gamma_{\rm cusp}^2 &=&
  \left(\frac{44 \pi ^4}{45}+\frac{88 \zeta _3}{3}-\frac{536      \pi    ^2}{27}
    +\frac{490}{3}\right) N_c^2
  + \left(-\frac{64 \zeta _3}{3}+\frac{80 \pi     ^2}{27}-\frac{1331}{27}\right) N_c n_f
  \nonumber\\&&\mbox{}
  -\frac{16 n_f^2}{27}
  \,,\nonumber\\
  \gamma_{\rm cusp}^3 &=&
  \left(-\frac{32 \pi ^4}{135}+\frac{1280 \zeta _3}{27}-\frac{304 \pi
      ^2}{243}+\frac{2119}{81}\right) N_c n_f^2+\left(\frac{128 \pi ^2
      \zeta    _3}{9} +224 \zeta _5 -\frac{44 \pi^4}{27} 
  \right.\nonumber\\&&\left.\mbox{}
    -\frac{16252 \zeta _3}{27} +\frac{13346 \pi ^2}{243}-\frac{39883}{81}\right) N_c^2    n_f+\left(\frac{64 \zeta
      _3}{27}-\frac{32}{81}\right)    n_f^3
  + \ldots
  \,.
\label{eq::gamma_cusp}
\end{eqnarray}
where the ellipses in $\gamma_{\rm cusp}^3$ indicate 
non-$n_f$ terms which are not yet known.
For $\gamma^{q}$ we have
\begin{eqnarray}
  \gamma_q^0 &=& -\frac{3 N_c}{2}
  \,,\nonumber\\
  \gamma_q^1 &=& 
   \left(\frac{\pi ^2}{6}+\frac{65}{54}\right) N_c n_f+\left(7 \zeta
    _3-\frac{5 \pi ^2}{12}-\frac{2003}{216}\right) N_c^2
  \,,\nonumber\\
  \gamma_q^2 &=& 
   \left(-\frac{\pi ^4}{135} -\frac{290 \zeta _3}{27} +\frac{2243 \pi
    ^2}{972}+\frac{45095}{5832}\right) N_c^2 n_f+\left(-\frac{4 \zeta
    _3}{27}-\frac{5 \pi ^2}{27}+\frac{2417}{1458}\right) N_c n_f^2
\nonumber\\&&\mbox{}
   +N_c^3 \left(-68 \zeta_5-\frac{22 \pi ^2 \zeta _3}{9} -\frac{11 \pi ^4}{54} +\frac{2107 \zeta
    _3}{18} -\frac{3985 \pi
    ^2}{1944}-\frac{204955}{5832}\right)
  \,,\nonumber\\
  \gamma_q^3 &=&
   N_c^3 \left[\left(-\frac{680 \zeta _3^2}{9} -\frac{1567 \pi
    ^6}{20412} +\frac{83 \pi ^2 \zeta
    _3}{9}  +\frac{557 \zeta _5}{9} +\frac{3557 \pi ^4}{19440} -\frac{94807 \zeta _3}{972} +\frac{354343 \pi
    ^2}{17496}
\right.\right.\nonumber\\&&\left.\left.\mbox{}
+\frac{145651}{1728}\right)
    n_f\right]+\left(-\frac{8
    \pi ^4}{1215} -\frac{356 \zeta _3}{243}-\frac{2 \pi ^2}{81}+\frac{18691}{13122}\right) N_c
    n_f^3+\left(-\frac{2}{3} \pi ^2 \zeta _3
\right.\nonumber\\&&\left.\mbox{}
+\frac{166 \zeta _5}{9}+\frac{331 \pi ^4}{2430} -\frac{2131 \zeta
    _3}{243} -\frac{68201 \pi
    ^2}{17496}-\frac{82181}{69984}\right) N_c^2 n_f^2
+ \ldots
  \,.
\label{eq::gamma_q}
\end{eqnarray}
For the finite part of $\log(F_q)$ we obtain
\begin{eqnarray}
\lefteqn{  \log(F_q)|^{(4)}_{\mbox{\tiny large-$N_c$, finite part}} = } \nonumber\\
&&
   \left(\frac{\pi ^2 \zeta _3}{27}-\frac{53 \zeta
    _5}{135}+\frac{761 \pi ^4}{7290} +\frac{52 \zeta _3}{243} +\frac{9883 \pi
    ^2}{4374}+\frac{1865531}{104976}\right) N_c n_f^3
    +\left(\frac{137 \zeta
    _3^2}{54}+\frac{1753 \pi ^6}{34020}
\right.\nonumber\\&&\left.\mbox{}
  +\frac{26 \pi ^2 \zeta _3}{81}  +\frac{1798 \zeta _5}{15} -\frac{58547
    \pi ^4}{58320} +\frac{386105 \zeta
    _3}{5832}-\frac{24172133 \pi
    ^2}{419904}-\frac{918437291}{1679616}\right) N_c^2 n_f^2
\nonumber\\&&\mbox{}
    +\left(  \frac{24427 \zeta _7}{144} + \frac{23 \pi ^2 \zeta
    _5}{108}    -\frac{1079 \pi ^4 \zeta _3}{3240}  + \frac{19705
    \zeta _3^2}{108}     +\frac{347
    \pi ^6}{9720}       -\frac{2509 \pi ^2 \zeta
    _3}{1296}-\frac{514217 \zeta _5}{720}
\right.\nonumber\\&&\left.\mbox{}
-\frac{10961 \pi ^4}{5832}  -\frac{11482507 \zeta _3}{5832}  +\frac{284977643 \pi
    ^2}{839808}+\frac{874566569}{209952}\right) N_c^3 n_f
    +\ldots\,,
\label{eq::logFq_fin}
\end{eqnarray}
The expressions in Eqs.~(\ref{eq::gamma_cusp}) and~(\ref{eq::gamma_q}) up
to three-loop order confirm the results in the
literature~\cite{Vogt:2000ci,Berger:2002sv,Moch:2004pa,Moch:2005tm,Baikov:2009bg,Becher:2009qa,Gehrmann:2010ue}
and the $N_c^3 n_{f}^3$ term of $\gamma_{\text{cusp}}^3$ agrees with the result of 
Ref.~\cite{Gracey:1994nn,Beneke:1995pq}. All other terms in the four-loop results
$\gamma_{\text{cusp}}^3$ and $\gamma_q^3$ and the finite part in
  Eq.~(\ref{eq::logFq_fin}) are new.


\section{Integrals with constant leading singularities}
\label{sect:weight}

Our calculation involves planar four-loop form-factor integrals. We
classified all such integrals and performed an integral reduction,
resulting in $99$ master integrals. Before discussing their
evaluation, we devote this section to our basis choice for these
integrals.

\subsection{Leading singularities and d-log forms}

In recent years it has become standard to use a basis, whenever
possible, of integrals having constant leading singularities. Leading
singularities \cite{Cachazo:2008vp} are essentially defined as multidimensional
residues of the Feyman loop integrand.

The usefulness of integrals with constant leading singularities was first
noticed in the context of maximally supersymmetric gauge theory, where the
answer appears to be naturally written in terms of them.  Building on
experience with such integrals in the literature, their systematic use was
advocated in Ref.~\cite{ArkaniHamed:2010gh}.  A particular highlight is an
all-$n$ expression, where $n$ is the number of external legs for the
integrand of two-loop maximally helicity violating amplitudes in
$\mathcal{N}=4$ super Yang-Mills theory.  In fact, it turns out that the
appearance of integrals with simple leading singularities in this theory is
very natural, as can be seen in the twistor approach
of~\cite{Lipstein:2012vs}, or when expressing leading singularities as certain
Grassmannian integrals~\cite{ArkaniHamed:2012nw}.  Although more examples are
known in the planar case, the concept of constant leading singularities also
carries over to the non-planar sector,
see~\cite{Gehrmann:2011xn,Arkani-Hamed:2014via,Bern:2014kca} for examples.

The use of such integrals is not limited to supersymmetric amplitudes,
as was pointed out in Ref.~\cite{Henn:2013pwa}.  Since then, they were applied to
countless calculations of scattering amplitudes required for phenomenology, see, 
e.g., Ref.~\cite{Caola:2014lpa}.  
Of course, more integrals are needed in QCD compared to supersymmetric theories.
In this context, it is perhaps interesting to point out that many of the
additional integrals needed for QCD can be thought of as integrals
being defined in a dimension shifted by two units. As is well known,
integrals in $D\pm 2$ and $D$ dimensions are related. The picture that
emerges is that one should not only classify integrals having constant
leading singularities in four dimensions, but in all integer (in
particular even) dimensions, and then relate them to the
four-dimensional case.

Let us give some one-loop examples of such integrals.  We define the
triangle integral near four dimensions
\begin{align}\label{one-loop-triangle}
  I_{\rm triangle} = \int \frac{{\rm d}^{4-2\eps }k}{i\pi^{2-\eps}}
  \frac{(p_1+p_2)^2}{k^2 (k+p_1)^2 (k-p_2)^2} \,, 
\end{align}
and the propagator-type integral near two dimensions,
\begin{align}\label{one-loop-bubble}
  I_{\rm bubble} = \int \frac{{\rm d}^{2-2\eps }k}{i\pi^{1-\eps}}
  \frac{(p_1+p_2)^2}{(k+p_1)^2 (k-p_2)^2} \,,
\end{align}
where $p_1^2=p_2^2=0$.

In the following we will consider leading singularities
at $\eps=0$.  It is convenient to change variables. For the bubble, we
set $k^\mu = \alpha p_1^\mu + \beta p_2^\mu$, which leads to
\begin{align}\label{bubble-param}
  \frac{{\rm d}^{2}k \,(p_1+p_2)^2}{(k+p_1)^2 (k-p_2)^2} \propto
  \frac{ {\rm d}\alpha\, {\rm d}\beta}{(\alpha+1)\alpha \beta (\beta-1)} \,,
\end{align}
where the proportionality sign means that the equation holds up
to kinematic-independent factors.
While there are various locations of the leading singularities, we can
see that all poles are kinematic-independent.  A similar analysis was
done for the triangle integral, see \cite{Bern:2014kca}.

We mention that one can rewrite the integrands (algebraically) in a
form where this property is manifest, namely,
\begin{align}\label{dlogbubble}
  \frac{ {\rm d}\alpha\, {\rm d}\beta}{(\alpha+1)\alpha \beta
    (\beta-1)} 
  =  \pm\,
  {\rm d}\log\left[ \frac{ (k+p_1)^2}{ (k-k_\pm)^2 }\right] {\rm d}\log\left[
    \frac{ (k-p_2)^2}{ (k-k_\pm)^2 }\right] \,. 
\end{align}
Here $k_{\pm}$ denotes the two solutions to the maximal cut condition,
$(k_{\pm}+p_1)^2=0, (k_\pm-p_2)^2=0$, which are given by $k_+=-p_1+p_2$
and $k_-=0$.  Equation~(\ref{dlogbubble})
implies that there exist variables in which the integrand is just ${\rm d}\log x_1
{\rm d}\log x_2$, with unit normalization.  More formulas of this type, called d-log
forms, can be found in Refs.~\cite{Arkani-Hamed:2014via,Bern:2014kca}.

\begin{figure}[t] 
  \begin{center}
    { \psfrag{p1}[cc][cc]{$p_{1}$}
      \psfrag{p2}[cc][cc]{$p_{2}$}
      \psfrag{k1}[cc][cc]{$k_{1}$}
      \psfrag{k2}[cc][cc]{$k_{2}$}
      \psfrag{k3}[cc][cc]{$k_{3}$}
      \psfrag{k4p1}[cc][cc]{$k_{4}+p_{1}$}
      \includegraphics[width=0.4\textwidth]{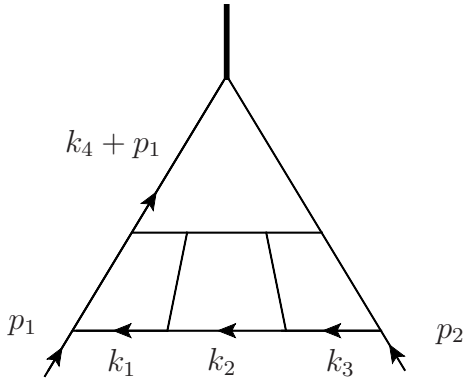}}
    \caption{Twelve-propagator form-factor integral that has unit leading
      singularities. The numerator $(k_{4}^2)^2$ normalization factor
      is implied.}
    \label{fig:12propagators}
  \end{center}
\end{figure}

Following these ideas, we wrote down a basis of integrals with
constant leading singularities for planar four-loop form-factor
integrals. The whole basis will be presented elsewhere.  Here we give
one example, for the twelve-propagator integral that was needed in the
$n_{f}$-calculation. See Fig.~\ref{fig:12propagators} and the first
diagram of Fig.~\ref{fig::diags} for a representative Feynman diagram.
We choose as basis element
\begin{align} \label{ff4l-ut-choice}
  I_{12 \;{\rm prop}} = \eps^8 (-s)^{1-4 \eps} e^{4 \eps \gamma_{\rm E}}
  & \int   \prod_{j=1}^{4}  \frac{ {\rm d}^{D}k_{j} }{i\pi^{D/2}} \,
  \frac{(k_{4}^2)^2}{k_{1}^2 k_{2}^2 k_{3}^2 (k_1 -k_2)^2 (k_2-k_3)^2
    (k_1-k_4)^2 }\nonumber \\  
  &\hspace{-2cm} \times \frac{1}{ (k_2-k_4)^2 (k_3-k_4)^2 (k_1+p_1)^2
    (k_4+p_{1})^2 (k_4-p_2)^2 (k_3-p_2)^2} 
\end{align}
The normalization factors were chosen for later convenience.  We first would
like to illustrate that this integral has indeed constant leading
singularities.  While this can be done algorithmically, it is instructive to
rewrite the integrand in a form where this is obvious, namely in terms of d-log
forms of the type discussed above.  A very useful feature is that this analysis can be
done loop by loop, which allows one to recycle formulas. This is very similar
to an analysis via cuts, although here we do not assume that any loop momenta
are on-shell.  First, we note that the 
box subintegrals with three off-shell legs, i.e. the ones depending on loop
momenta $k_1$ and $k_3$ (see Fig.~\ref{fig:12propagators}), can be written in
a d-log form.  For the subsequent calculation, only the normalization factor
of these subintegrals is relevant.  The latter can be obtained from any of
their leading singularities.  For example, for the box integral on the
left, we have the following integrand
\begin{align}
  \frac{{\rm d}^{4}k_{1}}{i\pi^2} \frac{1}{k_{1}^2 (k_1 + p_1)^2
    (k_1-k_4)^2 (k_1-k_2)^2}
  \,.
\end{align}
After taking a multi-dimensional residue one obtains either zero, or a
term proportional to
\begin{align}
  \frac{1}{k_4^2 (k_2+p_1)^2-k_2^2 (k_4+p_1)^2} 
  \,.
\end{align} 
For the box on the right a similar expression is obtained.
Next, we consider the $k_{2}$ integration.  Taking into account the factors
obtained from the $k_1$ and $k_3$ integrals, we arrive at a generalized box
integral
\begin{align}
  \frac{{\rm d}^{4}k_{2}}{k_{2}^2 (k_2 - k_4)^2  [k_4^2 (k_2+p_1)^2-k_2^2
      (k_4+p_1)^2]  [k_4^2 (k_2-p_2)^2-k_2^2  (k_4-p_2)^2] }\,. 
\end{align}
Again, it can be seen that this has a d-log form, with the
normalization factor $1/(k_{4}^2)^3$.  We now see that the numerator
in Eq.~(\ref{ff4l-ut-choice}) cancels the excessive factors of
$k_{4}^2$.  Indeed, putting everything together, we see that the
remaining $k_{4}$ integral is exactly of the form of the one-loop
triangle integral of Eq.~(\ref{one-loop-triangle}). In summary, this
proves that (\ref{ff4l-ut-choice}) has a d-log representation with
unit normalization.

We would like to emphasize again that the classification of integrals having
constant leading singularities can be done algorithmically.  Let us expand on
this point. First of all, for a given propagator structure, one makes an
ansatz for all possible numerator terms allowed by power counting (or subject
to other criteria).  It is convenient to parametrize the loop momentum in such
a way that the integration parameters are scalars. We illustrated this in the
case of the bubble integral, cf. Eq.~(\ref{bubble-param}).  Next, one
evaluates all leading singularities of this ansatz (i.e., one computes the
residues at all poles of the integrand). Requiring that the residues be
kinematic-independent yields a system of equations, which is then solved. It
is important to realize that this analysis only depends on the integrand at
hand, and can be done before attempting to compute the integral.

\subsection{Transcendental weight properties}

One nice feature of integrals with constant leading singularities is
that, conjecturally, they evaluate to so-called pure functions,
i.e. iterated integrals of uniform weight.

For iterated integrals, such as multiple polylogarithms, the weight is defined
as the number of integrations, e.g. one for logarithms, $n$ for classical
polylogarithms ${\rm Li}_{n}$, etc.  Similarly, transcendental constants such
as zeta values, $\zeta_n$, have weight $n$.  Finally, when considering Laurent
expansions in the dimensional regularization parameter $\eps$, one can assign
weight $-1$ to $\eps$. This is natural since $1/\eps$ would be represented by
logarithm in a cutoff regularization.

With these definitions, we see that the triangle integral of
Eq.~(\ref{one-loop-triangle}) has uniform weight $2$,
\begin{align}
  (-s)^{1-\eps} e^{\eps \gamma_{\rm E}} I_{\rm triangle} =
  \frac{1}{\eps^2} -\frac{1}{12} \pi^2 -\frac{7}{3} \zeta_3 \eps -
  \frac{47}{1440} \pi^4 \eps^2 + \mathcal{O}(\eps^3)  \,. 
\end{align}
More generally, $L$-loop integrals with constant leading singularities
in $4$ dimensions are expected to evaluate to weight $2L$ functions.

Beyond maximally supersymmetric Yang-Mills theory, also functions of weight
smaller than $2L$ are needed. Perhaps the best way to understand the
additional integrals is to consider them in different dimensions. 
Take as an example the $2-2 \eps$ dimensional bubble integral of
Eq.~(\ref{one-loop-bubble}).  In fact, in this simple example, the bubble and
triangle integrals are related by an integration-by-parts relation (and,
dimensional shift relation~\cite{Tarasov:1996br,Lee:2009dh}), which implies
\begin{align}
  I_{\rm bubble} = -2 \eps\, I_{\rm triangle}   \,,
\end{align}
where the integrals are defined in Eqs.~(\ref{one-loop-triangle})
and~(\ref{one-loop-bubble}).
From this formula we see that its weight is shifted by one compared to the triangle. 
It evaluates to a uniform weight-one function.

More generally, at higher loops one can also generate integrals of
various weights, in particular by writing subintegrals formally in
different dimensions. For example, all the uniform weight integrals
presented in Ref.~\cite{Henn:2013pwa} and elsewhere can be understood
in this way.  See also the lecture notes \cite{Henn:2014qga} for more
details.

Returning to our form-factor integrals, we can verify the uniform
weight conjecture for the most complicated twelve-propagator integral of
Eq.~(\ref{ff4l-ut-choice}).  As a result of the calculation of the
next section, we find
\begin{align} \label{ff4l-ut-choice-result}
  I_{12 \;{\rm prop}} =& \frac{1}{576} + \eps^2 \frac{1}{216}\pi^2 +
  \eps^3 \frac{151}{864} \zeta_{3} + \eps^4 \frac{173 }{10368}\pi^4 +
  \eps^5 \left[ \frac{505}{1296} \pi^2 \zeta_{3} +
    \frac{5503}{1440}\zeta_{5} \right]+ \nonumber \\ &\hspace{-0cm} +
  \eps^6 \left[ \frac{6317 }{155520}\pi^6 + \frac{9895
    }{2592}\zeta_{3}^2 \right] + \eps^7 \left[ \frac{89593
    }{77760}\pi^4 \zeta_{3} + \frac{3419}{270} \pi^2 \zeta_{5} -
    \frac{ 169789 }{4032}\zeta_{7} \right] \nonumber \\ &\hspace{-0cm}
  + \eps^8 \left[ \frac{407}{15} s_{8a} + \frac{41820167}{653184000}
    \pi^8 + \frac{41719}{972} \pi^2 \zeta_{3}^2 - \frac{263897}{2160}
    \zeta_{3} \zeta_{5} \right] +\mathcal{O}(\eps^9) \,,
\end{align}
where $s_{8a} = \sum_{i_1 = 1}^{\infty} \frac{1}{i_1^5}
\sum_{i_2=1}^{i_1} \frac{1}{i_2^3} =\zeta_{8} + \zeta_{5,3} =
1.041785...$ and $\zeta_{5,3}$ is a multiple zeta value
\cite{Maitre:2005uu}.  Reinstating the $1/\eps^8$ from
Eq.~(\ref{ff4l-ut-choice}), one sees that this is a uniform weight
eight integral, as expected from a four-loop integral.  As an
independent check of Eq.~(\ref{ff4l-ut-choice-result}), we derived a
Mellin-Barnes representation (see Chapter~5 of \cite{Smirnov:2012gma} 
for a review) for this integral, which we used to
verify the first three terms in the $\eps$ expansion analytically.


\section{\label{sec::dgl}Differential equation bootstrap for single-scale integrals}

\begin{figure}[t]
  \psfrag{p1}[cc][cc]{$p_{1}$}
  \psfrag{p2}[cc][cc]{$p_{2}$}
  \psfrag{p3}[cc][cc]{$p_{3}$}
  \psfrag{x=0}[cc][cc]{$x = 0$}
  \psfrag{x=1}[cc][cc]{$x = 1$}
  \psfrag{Re(x)}[cc][cc]{${\rm Re}(x)$}
  \psfrag{Im(x)}[cc][cc]{${\rm Im}(x)$}
  \begin{center}\subfloat[]{\includegraphics[width=0.4\textwidth]{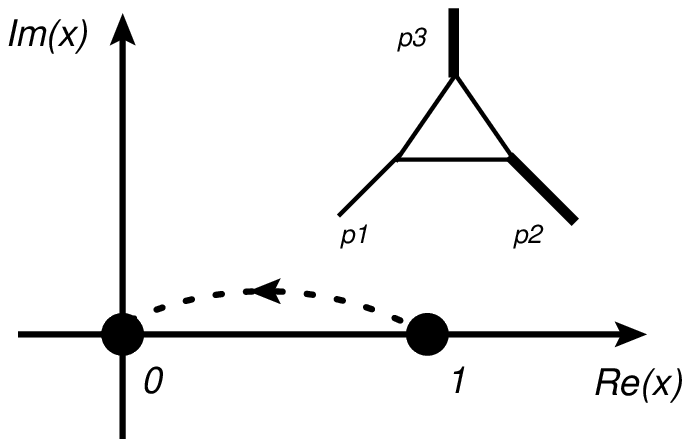}}
    \quad
    \subfloat[]{\includegraphics[width=0.2\textwidth]{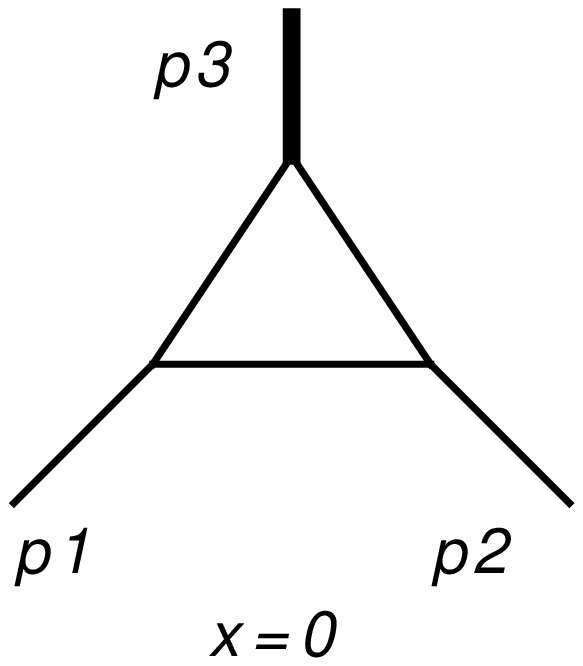}}
    \quad
    \subfloat[]{\includegraphics[width=0.23\textwidth]{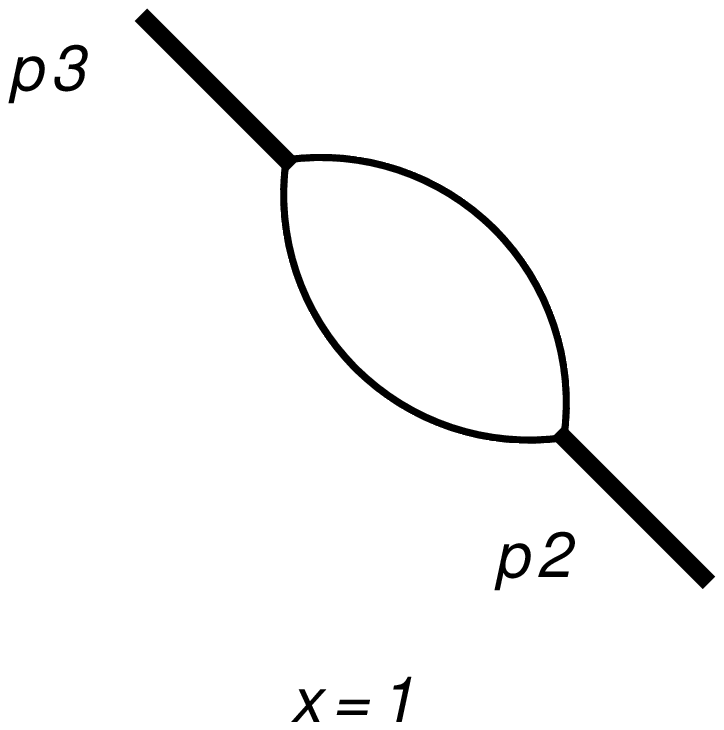}}
    \caption{Bootstrapping on-shell form-factor integrals at $x=0$ (b)
      from propagator integrals at $x=1$ (c).  The
      form factor with two off-shell legs is shown in (a), where $x=p_2^2/p_3^2$.}
    \label{fig:bootstrap}
  \end{center}
\end{figure}

In this section we discuss the first analytic computation of all planar
four-loop on-shell form-factor integrals which are defined in the kinematic
regime $p_{1}^2 = p_2^2 = 0$, with ${q^2\equiv p_3^2}=(p_1 +p_2)^2$.
Following the strategy of~\cite{Henn:2013nsa} we introduce an auxiliary
parameter by taking a second external leg off-shell, i.e.  $p_{2}^2 \neq 0$,
and $x=p_{2}^2/{p_3^2}$, and derive differential equations with respect to $x$.  The
main idea can be explained via Fig.~\ref{fig:bootstrap}.  It turns out that
the singular points of the differential equations are $x=0,1,\infty$.  The
point $x=0$ corresponds to the original on-shell form-factor integrals,
Fig.~\ref{fig:bootstrap}(b). 
Similarly, for $x=\infty$ we have $p_3^2=0$, which again leads to form-factor integrals,
as for $x=0$, and thus this case does not have to be considered separately.
On the other hand, the boundary value at $x=1$
corresponds to propagator-type integrals, see Fig.~\ref{fig:bootstrap}(c).
They can be determined easily: in most cases, the boundary value is zero due to
kinematical factors. Otherwise one can use results for propagator type
integrals available in the literature, see, in particular, four-loop analytic
results in \cite{Baikov:2010hf,Lee:2011jt,Panzer:2015ida}.  This information
is then transported back via the differential equation to $x=0$, see
Fig.~\ref{fig:bootstrap}(a).  Let us now see how this works in a bit more
detail.  A pedagogical example is given in \cite{Henn:2014qga}.

In reference \cite{Henn:2013pwa}, a canonical form of differential
equations for Feynman integrals was suggested. Conjecturally, this
form can be reached whenever the master integrals can be chosen to
have the property that their leading singularities are constant, as
explained in Section~\ref{sect:weight}.  This reduces the problem of
finding a canonical basis for the differential equations to a simple
classification of integrals having this property. The latter can be
done algorithmically.

For the planar form factor with $p_2^2\not=0$ and $p_3^2\not=0$
we find a total of $504$ master integrals (some
of them related by symmetry). After choosing a canonical basis
$\vec{f}$, we found the following system of differential equations,
\begin{align}\label{de504}
  \partial_x \vec{f}(x,\epsilon) = \epsilon \left[ \frac{a}{x} +
    \frac{b}{1-x} \right] \vec{f}(x,\epsilon)\,,
\end{align}
where $a$ and $b$ are some constant (i.e. $x$- and $\eps$-independent)
$504\times 504$ matrices.  The special features of this form are the manifest
Fuchsian property of the singularities, i.e. only single poles in
$x=0,1,\infty$ are present on the right-hand side of Eq.~(\ref{de504}),
and the fact that the right-hand side is proportional to $\epsilon$.
The latter property can be achieved for iterated integrals. Here, it implies
that the solution, to any order in $\epsilon$, can be written in terms of 
iterated integrals over the kernels ${\rm d}x/x$ and ${\rm d}x/(x-1)$, i.e. in
terms of harmonic polylogarithms \cite{Remiddi:1999ew}.
The former property is true for any Feynman integral. Making it
manifest allows us to describe the boundary behavior in a simple way,
namely
\begin{align}
  \vec{f}(x,\epsilon) \stackrel{x\to 0}{=} & 
  \left[ 1 + \sum_{k \ge
      1} p_{k}(\epsilon) x^k \right] x^{\epsilon a}
  \vec{f}_{0}(\epsilon) \,, \\ 
  \vec{f}(x,\epsilon) \stackrel{x\to 1}{=} & 
  \left[ 1 + \sum_{k \ge 1} q_{k}(\epsilon) (1-x)^k
    \right] (1-x)^{{-\epsilon b}} \vec{f}_{1}(\epsilon) \,,
\end{align}
where $a$ and $b$ are the matrices from Eq.~(\ref{de504}) and the coefficients
matrices $p_k(\epsilon)$ and $q_{k}(\epsilon)$ in the expansion can be
obtained recursively~\cite{Wasow}.

We fix the
boundary value at $x=1$ by demanding regularity of the integrals in this limit
and using explicit results for some propagator type integrals. 
This determines $\vec{f}_{1}(\epsilon)$.

We then use the differential equation (\ref{de504}) to transport this boundary
value back to $x=0$.  (In mathematical language, we construct the Drinfeld
associator, perturbatively in $\epsilon$.)  This allows us to determine
$\vec{f}_{0}(\epsilon)$.  Finally, unlike the $x\to 1$ limit, the $x\to 0$
limit is singular, in the sense that the matrix exponential $x^{\epsilon a}$
contains several terms $x^{\epsilon \alpha}$, with $\alpha \neq 0$.  These
non-zero values of $\alpha$ correspond to contributions of various regions
\cite{Beneke:1997zp,Smirnov:1998vk,Smirnov:2002pj} to the asymptotic expansion
in the given limit.  The on-shell integrals we would like to compute
correspond to the so-called ``hard'' region with $\alpha =0$.

In order to determine to the on-shell integrals, we reduce the basis $\vec{f}$
for on-shell kinematics, expressing it in terms of $99$ on-shell master
integrals. We then match the expression so obtained to the hard region at
$x=0$. We find that this determines all the $99$ integrals (naturally, some of
the 504 equations are redundant).  In order to carry out these algebraic
manipulations, we found the {\rm Mathematica} implementation {\tt
  HPL.m}~\cite{Maitre:2005uu} useful.

In summary, we analytically computed all planar form-factor integrals with two
off-shell legs (504 master integrals), and with one off-shell leg (99 master
integrals). The result for the most complicated on-shell form-factor integral
with twelve propagators that appeared in the $n_{f}$-piece of our calculation
was given in eq. (\ref{ff4l-ut-choice-result}) as an example. The full
analytic results for all planar master integrals will be given elsewhere.


\section{\label{sec::concl}Conclusions}

In this paper we report the calculation of all 504 master integrals
which are needed for a generic planar massless form factor with two off-shell
legs. They are obtained by a proper choice of basis integrals, together with
boundary conditions where the form factor degenerates to a two-point function.
From the generic basis we derive analytic results for the 99 master integrals
that are needed for the planar on-shell form factor. 78 out of the 99 master
integrals are needed for the fermionic part of the planar photon-quark form
factor. The latter is considered in Section~\ref{sec::calc_res} of this paper,
where analytic results up to four loops are presented for the cusp and
collinear anomalous dimension and the finite part of $F_q$.

A natural extension of this work is to apply the planar master
integrals we computed to evaluate the non-fermionic planar
contribution, where the integral reduction is more complicated.
Furthermore,
we expect that the methods discussed in this paper can also be applied to
non-planar form factor integrals.


\section*{Acknowledgments}

J.M.H. and M.S. acknowledge support via the DFG project ``Infrared and
threshold effects in QCD''. V.S. is grateful to Gang Yang for comparison of 
some of our analytical results with numerical results of \cite{Boels:2015yna}.
We thank Andreas Vogt for discussions on the constant $n_f^2$
term of $\gamma_{\rm cusp}^3$ in Eq.~(\ref{eq::gamma_cusp}) which helped us to fix
a typo in Eq.~(\ref{eq::logFq_gen})  in the first version of this paper.


\bibliographystyle{JHEP} 

\bibliography{ff4l_nf_bib}


\end{document}